\documentclass[10pt]{article}
\usepackage{graphicx}

\def\Title#1{\begin{center} {\Large #1 } \end{center}}
\def\Author#1{\begin{center}{ \sc #1} \end{center}}
\def\Address#1{\begin{center}{ \it #1} \end{center}}

\newcommand\pubblock{\rightline{\begin{tabular}{l} Proceedings of the Fifth Annual LHCP\\ \pubnumber\\
         \pubdate  \end{tabular}}}

\newenvironment{Abstract}{\begin{quotation} \begin{center} 
             \large ABSTRACT \end{center}\bigskip 
      \begin{center}\begin{large}}{\end{large}\end{center} \end{quotation}}

\newenvironment{Presented}{\begin{quotation} \begin{center} 
             PRESENTED AT\end{center}\bigskip 
      \begin{center}\begin{large}}{\end{large}\end{center} \end{quotation}}

\def\beq{\begin{equation}}
\def\eeq#1{\label{#1}\end{equation}}
\def\eeqn{\end{equation}}

\def\beqa{\begin{eqnarray}}
\def\eeqa#1{\label{#1}\end{eqnarray}}
\def\eeqan{\end{eqnarray}}

\let\bar=\overbar

\def\Dslash{\not{\hbox{\kern-4pt $D$}}}
\def\dslash{\not{\hbox{\kern-2pt $\del$}}}

\def\msb{{\bar{\ssstyle M \kern -1pt S}}}

\usepackage{color}

\newcommand\pubnumber{ CMS-PAS-SUS-17-039 }

\newcommand\pubdate{\today}

\def\affiliation{
Universidad de Oviedo}

\begin{document}

\large
\begin{titlepage}
\pubblock

\vfill
\Title{ Search for electroweak production SUSY in same-sign dileptons and multilepton final states with CMS }
\vfill

\Author{ Juan R. Gonz{\'a}lez Fern{\'a}ndez \\ on behalf of the CMS Collaboration}
\Address{\affiliation}
\vfill
\begin{Abstract}
Searches for direct electroweak production of charginos and neutralinos in signatures with two light leptons of the same charge and with three or more leptons including up to two hadronically decaying $\tau$ leptons are presented . The full 2016 dataset of pp collisions recorded by CMS at a center-of-mass energy of 13 TeV is used, corresponding to an integrated luminosity of 35.9 $fb^{-1}$. The observed event rates are in agreement with expectations from the standard model. These results probe charginos and neutralinos with masses up to values between 225 and 1150 GeV, depending on the model parameters assumed.
\end{Abstract}
\vfill

\begin{Presented}
The Fifth Annual Conference\\
 on Large Hadron Collider Physics \\
Shanghai Jiao Tong University, Shanghai, China\\ 
May 15-20, 2017
\end{Presented}
\vfill
\end{titlepage}
\def\thefootnote{\fnsymbol{footnote}}
\setcounter{footnote}{0}
%

\normalsize 


\section{Introduction and models}
The Standard Model (SM) describes very accurately most of the phenomena in particle physics. So far, searches for physics beyond the SM have not revealed convincing evidence for the existence of such phenomena. However, there are several open questions not covered by the SM, such as the hierarchy problem. Supersymmetry (SUSY) is an extension of the SM that can give an answer to that questions.

\begin{figure}[htb]
\centering
\includegraphics[height=2cm]{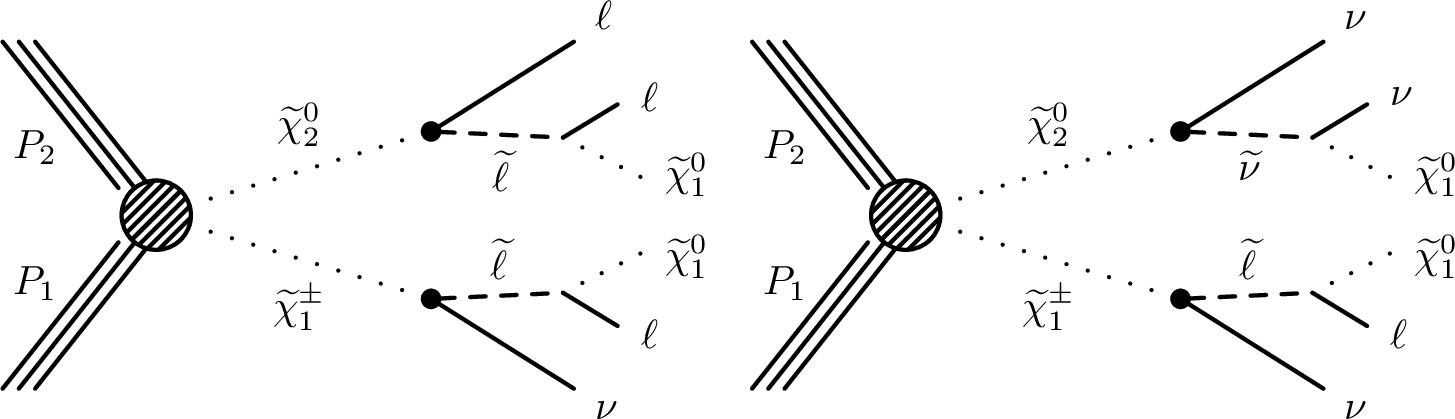}
\caption{Chargino and neutralino pair production with decays mediated by sleptons and sneutrinos.}
\label{fig:pro1}
\end{figure}

\begin{figure}[htb]
\centering
\includegraphics[height=2cm]{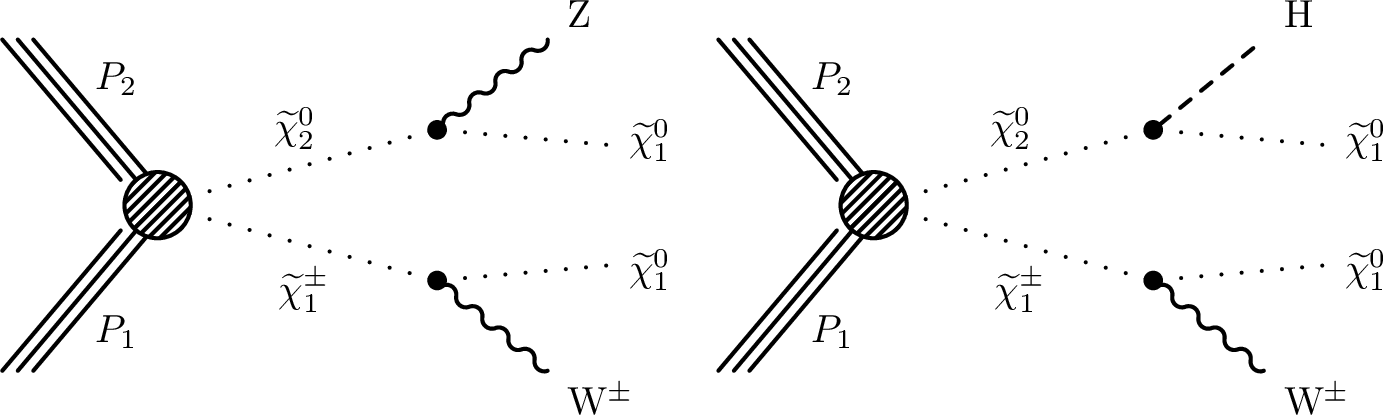}
\caption{Chargino and neutralino pair production with the chargino decaying to a W boson and the LSP and the neutralino decaying to a Z boson and the LSP (left) or a Higgs boson and the LSP (right).}
\label{fig:pro2}
\end{figure}

So far, no evidence for supersymmetry has been found and direct searches in accelerators have placed many constraints. In this comunication, the recent searches for electroweak SUSY particles in multilepton final states in CMS experiment \cite{bib:SUS16039} are shown. In this research, we focus on electroweak production of charginos and neutralinos decaying into leptons. In Figure \ref{fig:pro1} and \ref{fig:pro2} the main targeted production models are shown.

\section{Search strategy}
For this study, data are collected with CMS detector \cite{bib:cms}, recording a dataset with a total luminosity of 35.9 $fb^{-1}$. All particles are reconstructed using the Particle Flow algorithm \cite{bib:pf}. We select events with at least two light leptons (e, $\mu$) or one light lepton and two hadronically decaying tau leptons ($\tau_{h}$). To record these events, dileptonic and single lepton triggers are used.

Electrons (muons) are required to have $\vert \eta \vert <$ 2.5 (2.4) and $p_{T} >$ 10 GeV. Selected light leptons must also satisfy certain isolation criteria and have an transverse $d_0$ (longitudinal $d_z$) impact parameter that not exeed 0.5 (0.1) mm. Hadronic taus must have $\vert \eta \vert <$ 2.3 and $p_{T} >$ 20 GeV. Jets must have $\vert \eta \vert <$ 2.4 and $p_{T} >$ 25 GeV and must have an angular separation with respect to the light leptons of $\Delta R >$ 0.4. Selected jets are considered as comming from b quarks (b jets) using an algorithm with a efficiency of about 63\% and a mistagging rate of 1.5\% for light-flavor jets.

To optimize the sensitivity to the target models, events are clasified in multiple signal regions. First, events are clasified in categories with two same-sign leptons, three leptons and four or more leptons. For the categories with 3 or more leptons, an additional classification in number of $\tau_{h}$ leptons and existence of opposite-sign same-flavor lepton pairs is done. Further binning is done in sensitive variables such as missing transverse energy (MET), dilepton $p_T$, tansverse mass ($M_T$) or invariant mass ($M_{\ell\ell}$).

\section{Background estimate}
Some SM processes lead to similar final states. These processes will be considered backgrounds and need to be estimated. Backgrounds will be divided in the next categories:
\begin{itemize}

\item WZ and W$\gamma*$ production: When both bosons decay leptonically we obtain a targeted final state. This background is dominant in the searches with three light leptons and also for searches with two same-sign leptons, if a third lepton is out of the detector acceptance. This background is estimated from MC and normalized to data in a dedicated control region. MET and $M_T$ shapes are validated in data.

\item Nonprompt e, $\mu$, $\tau_h$: 
Instrumental background dominated by W+jets, $t\bar{t}$ or Drell-Yan processes. This background is dominant in the same-sign regions and some regions with 3 leptons. It is estimated from data, using a fake-rate measurement.

\item External and internal conversions: production of W or Z bosons or $t\bar{t}$ accompanied by radiation which suffers an asymmetric conversion. The low-$p_T$ lepton may be lost and the process can show a final state similar to the signal in some regions, contributing mainly to same-sign dilepton signal regions. This background is estimated from MC simulation and modeling checked in dedicated data control region.

\item Rare SM processes: including multiboson production and top-pair production in asociation with a boson. The contribution of such processes is estimated from MC simulation.

\item Electron charge misidentification: this background affects mainly to same-sign dilepton regions, when the charge of an electron is mismeasured in events with two opposite-sign leptons. In most cases, this arises from severe bremsstrahlung in the tracker material. This background is estimated by reweighting data events with two opposite-sign leptons by charge misidentification probability.

\end{itemize}

\section{Systematic uncertainties}
The considered systematic uncertainties are summarized in Table \ref{tab:syst}.

\begin{table}[h!]
\begin{center}
\small
\begin{tabular}{l|cc}
\hline
Source          & Estimated uncertainty (\%) & Treatment \\
\hline
e/$\mu$ selection &  3  & normalization \\
$\tau_{h}$ selection &  5  & normalization \\
Trigger efficiency &  3 & normalization \\
Jet energy scale & 2--10 & shape \\
b tag veto & 1--2 & shape \\
Pileup & 1--5 & shape\\
Integrated luminosity & 2.5 & normalization \\
\hline
Scale variations and PDF ($t\bar{t}Z$ and $t\bar{t}W$) & 15 & normalization \\
Theoretical (ZZ) & 25 & normalization \\
Conversions & 15 & normalization \\
Other backgrounds & 50 & normalization \\
Monte Carlo statistical precision & 1--30 & normalization \\
\hline
Nonprompt leptons (closure) & 30 & normalization \\
Nonprompt leptons (W/Z bkg. subtraction) & 5--20 & shape \\
Charge misidentification & 20 & normalization \\
WZ normalization & 10 & normalization\\
WZ shape & 5--50 & shape\\
\hline
ISR uncertainty                           & 1--5 & shape  \\
Scale variations for signal processes     & 1--2 & shape  \\
Lepton efficiencies       & 2 & normalization \\
Signal acceptance (METmodeling) & 1--5 & shape \\
\hline
\end{tabular}
    \caption{Summary of systematic uncertainties in the event yields in the search regions and their treatment.} 
    \label{tab:syst}
\end{center}
\end{table}

\section{Results and interpretations}
\begin{figure}[!hbtp]
\centering
\includegraphics[width=0.25\textwidth]{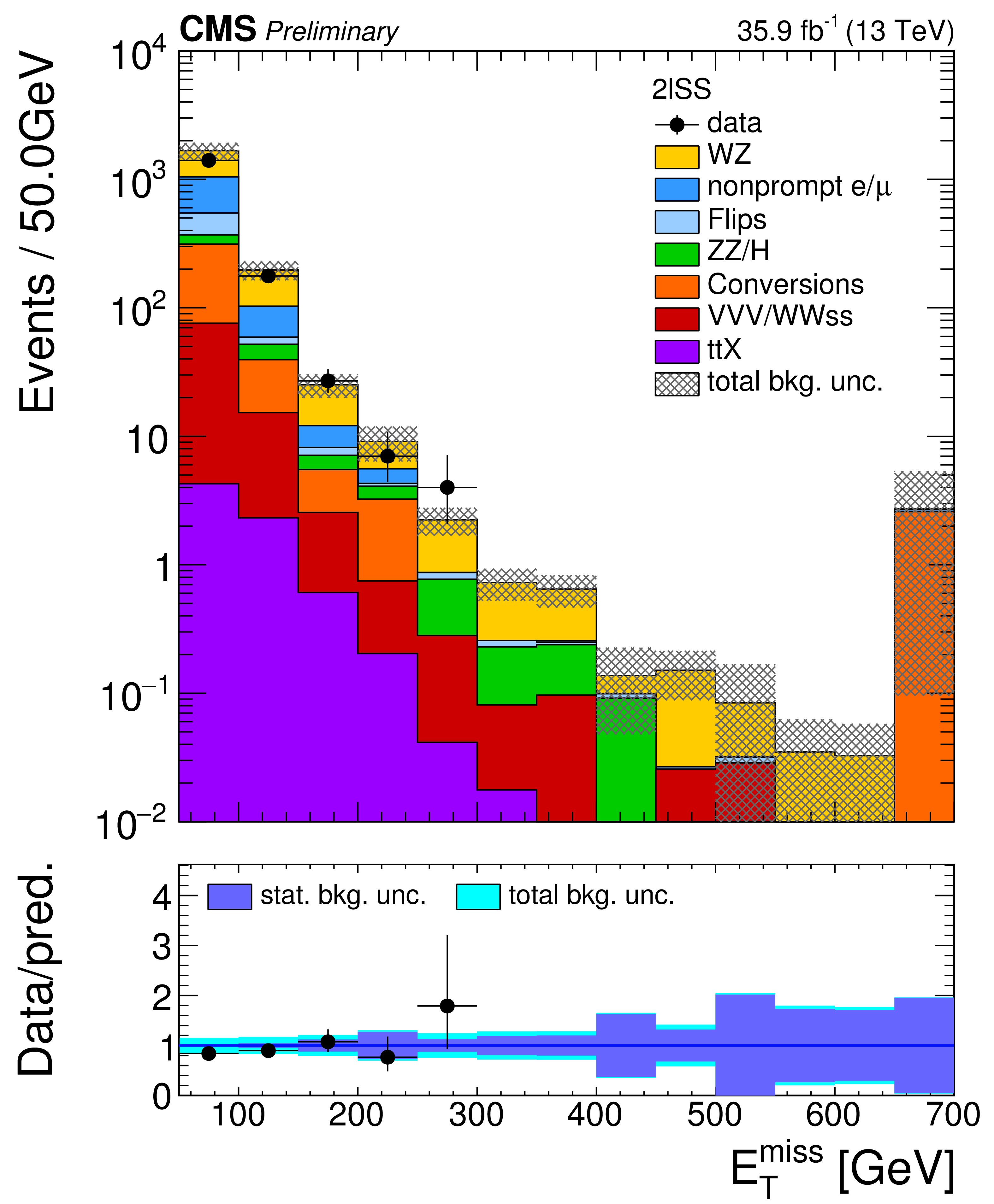}
\includegraphics[width=0.25\textwidth]{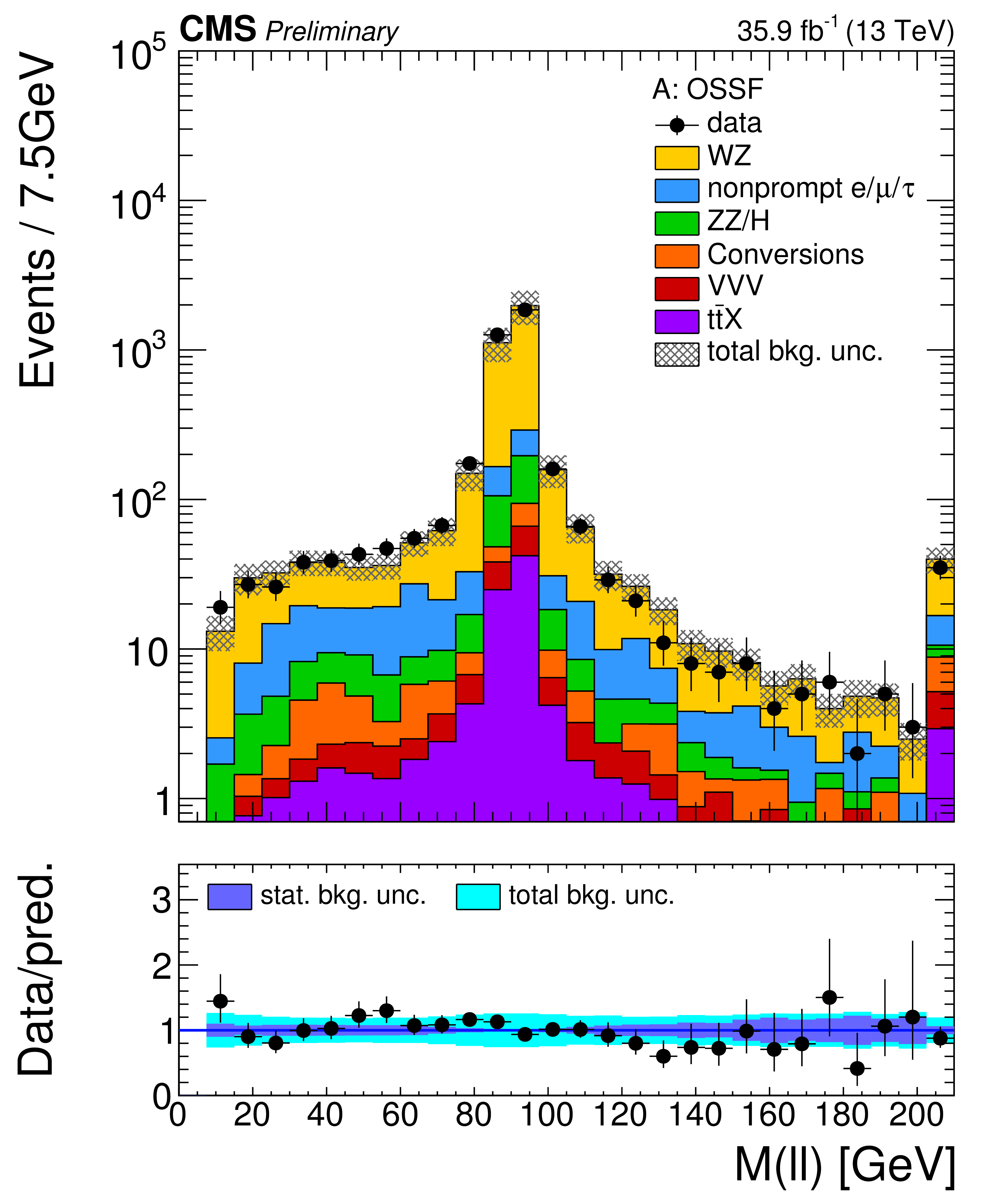} \\
\includegraphics[width=0.25\textwidth]{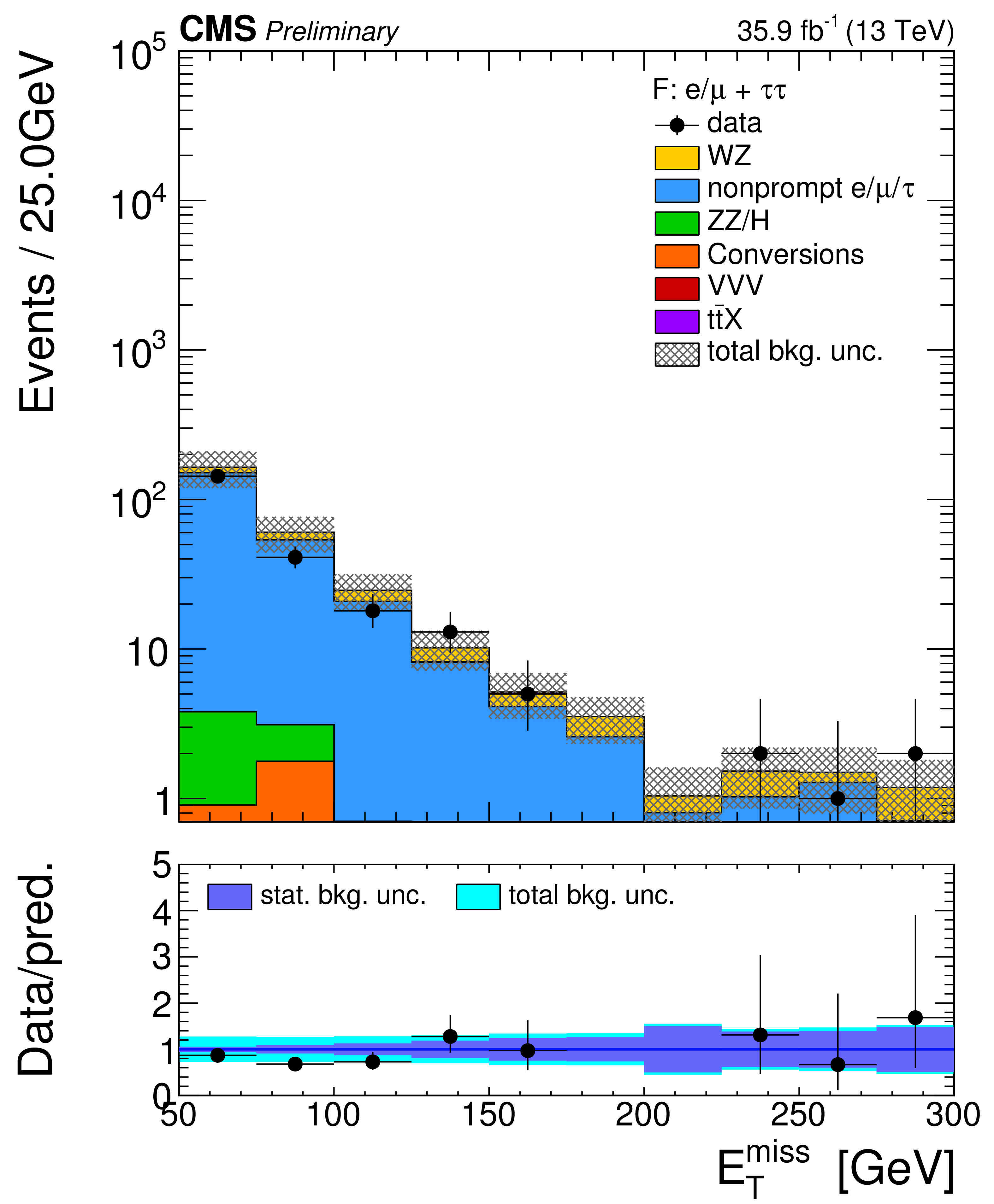}
\includegraphics[width=0.25\textwidth]{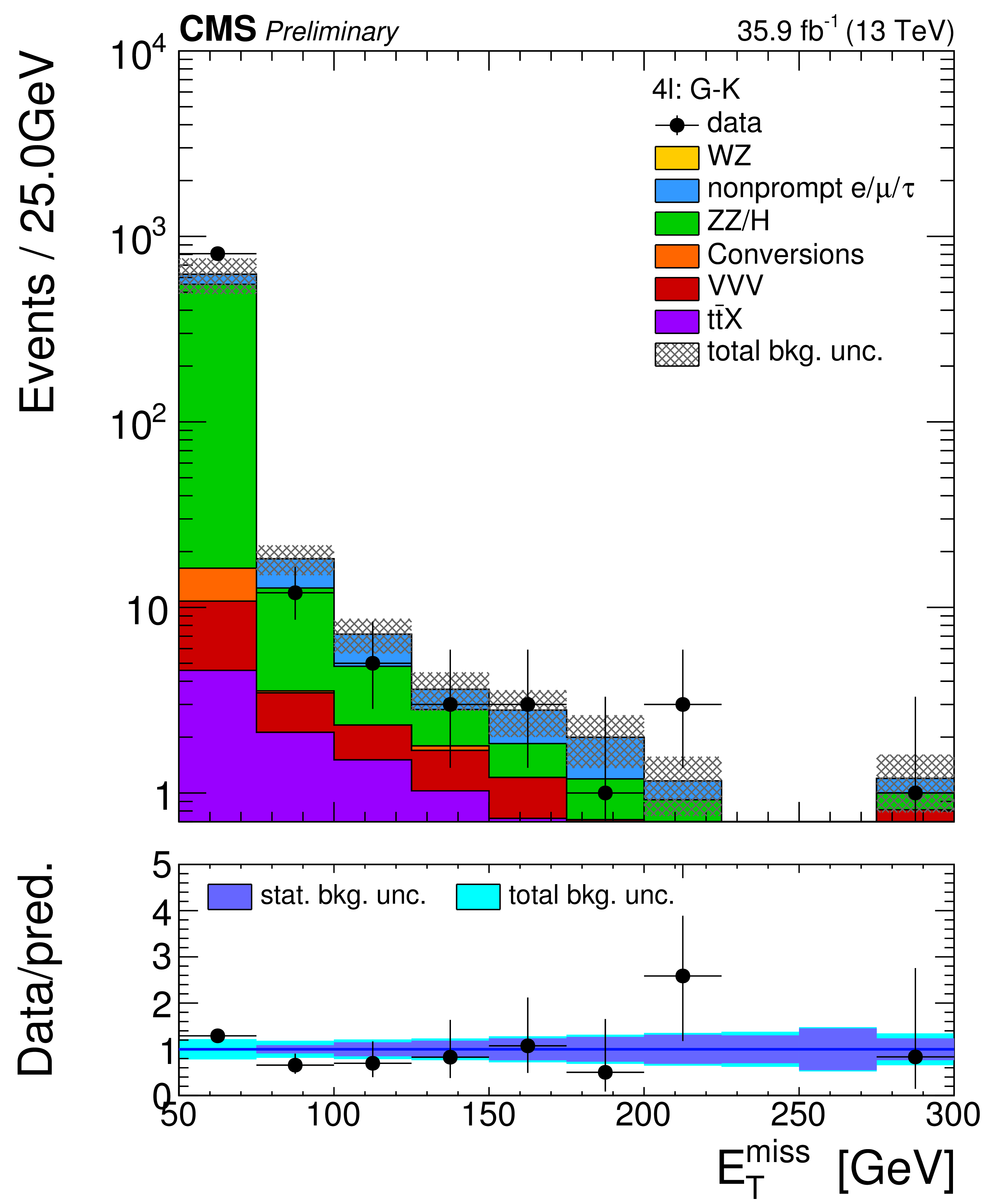}
\caption{Kinematic distributions of MET for events with 2 same-sign leptons and no jets (top left), invariant mass of the opposite-sign same-flavour pair for events with three leptons (top right) and MET for events with one light lepton and two $\tau_{h}$ (botom left) and four light leptons (botom right).}
\label{fig:kin}
\end{figure}

Distribution of representative kinematic distributions are shown in Figure \ref{fig:kin} 
for different event selections.

\begin{figure}[!hbtp]
\centering
\includegraphics[width=0.25\textwidth]{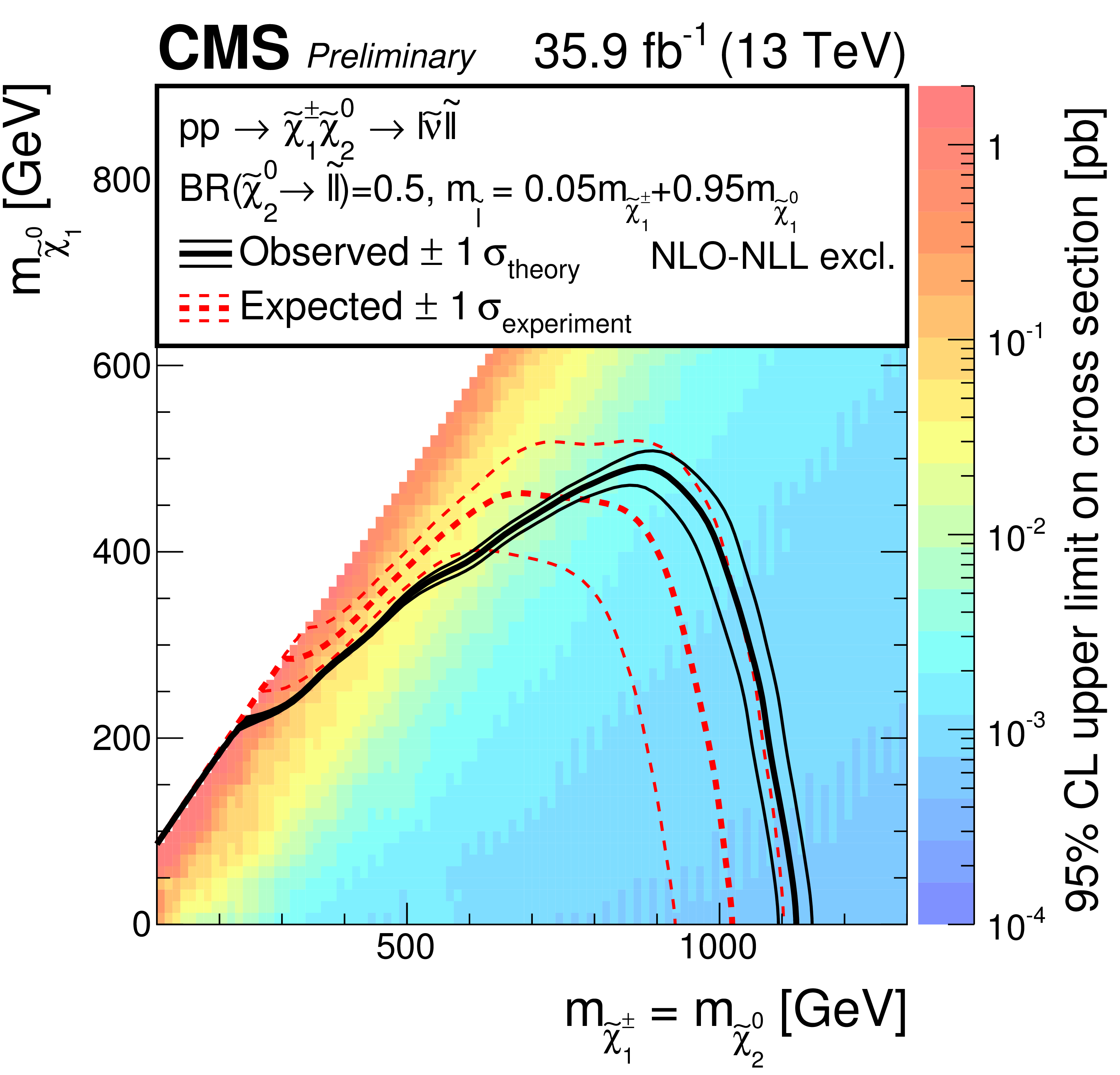}
\includegraphics[width=0.25\textwidth]{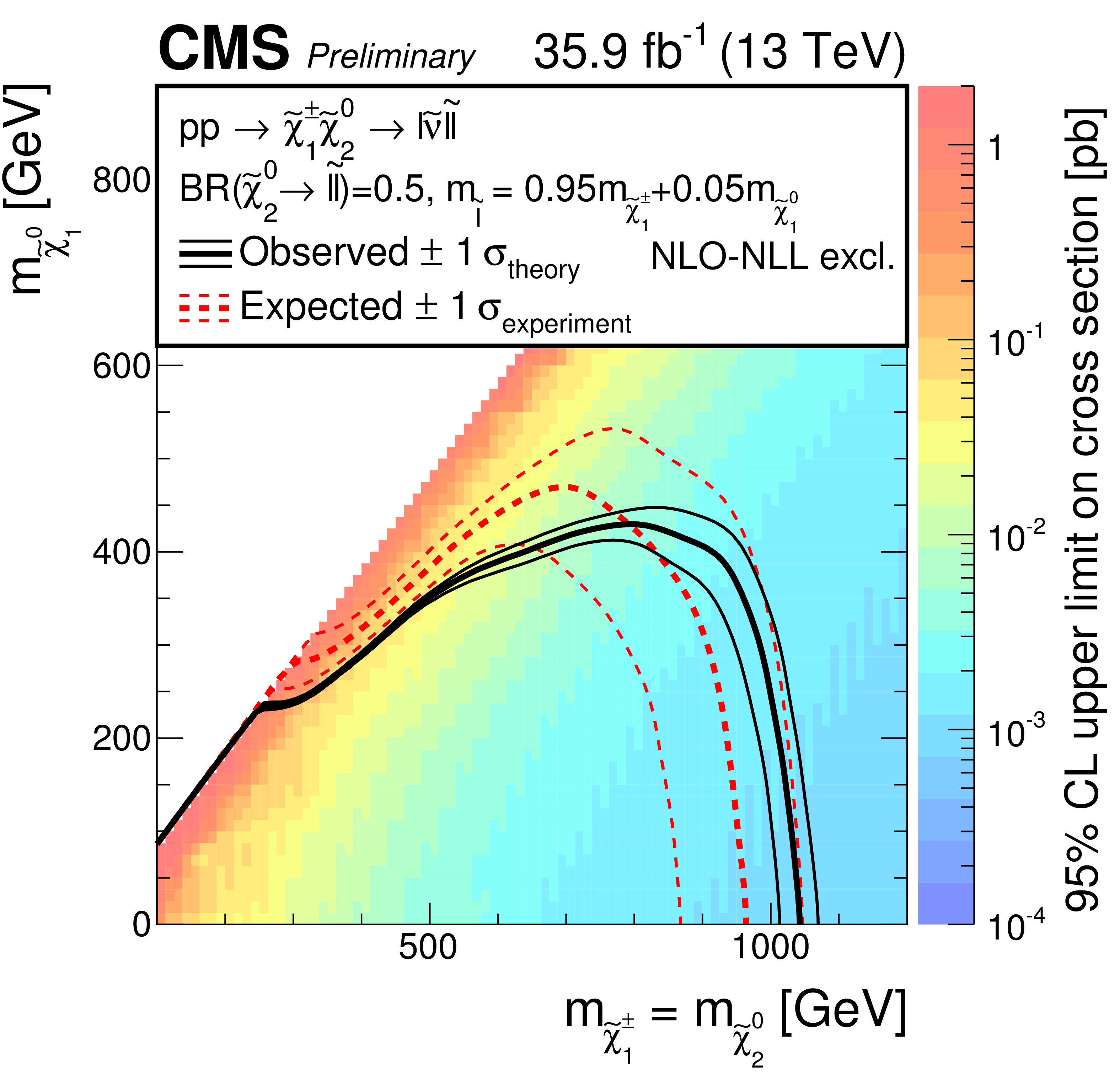} 
\caption{Interpretation for production of charginos and neutralinos in multilepton final 
states for flavour democratic model with mass parameter x = 0.05 (left) and x = 0.95 (right). }
\label{fig:int1}
\end{figure}

\begin{figure}[!hbtp]
\centering
\includegraphics[width=0.25\textwidth]{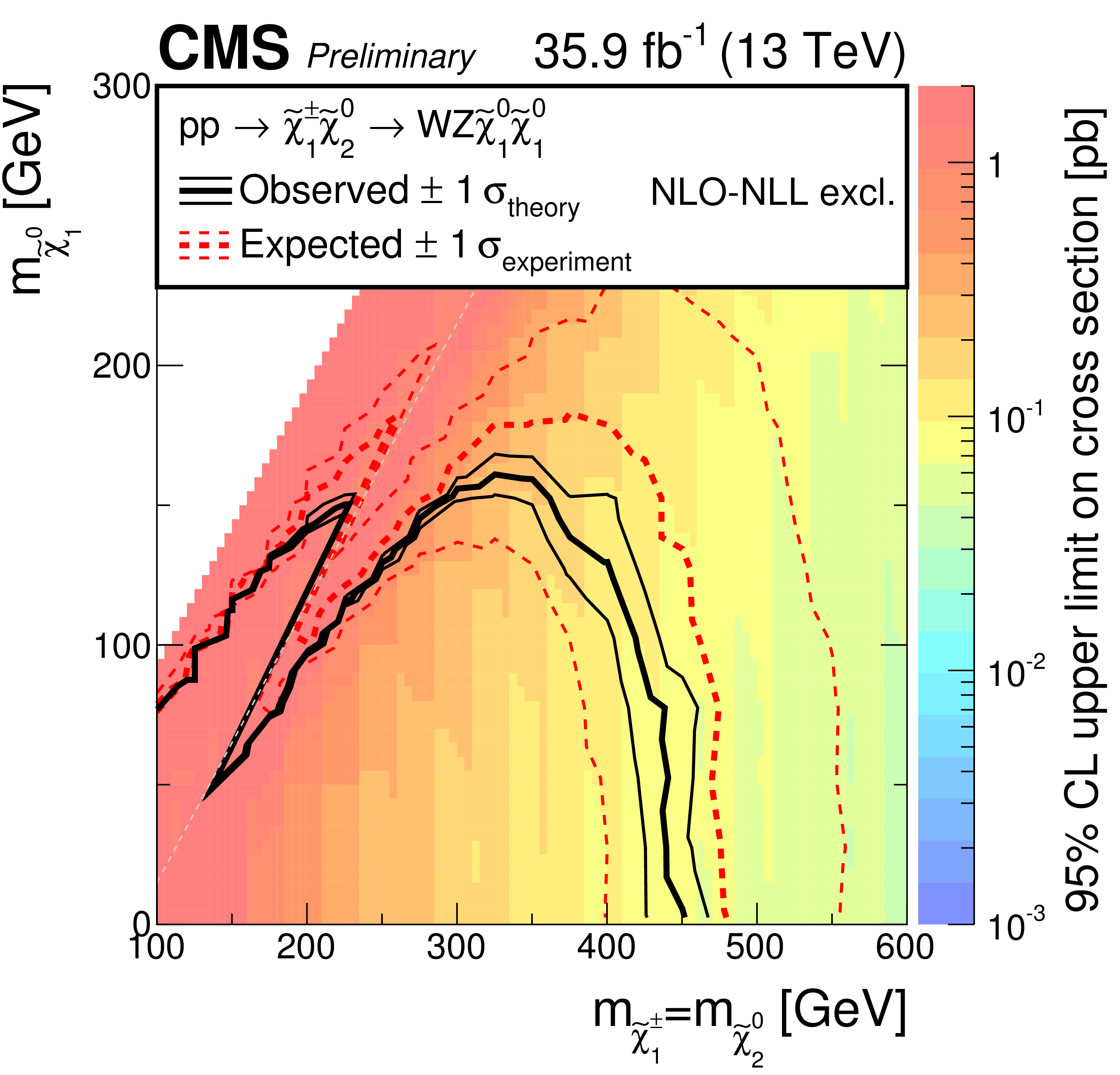}
\includegraphics[width=0.25\textwidth]{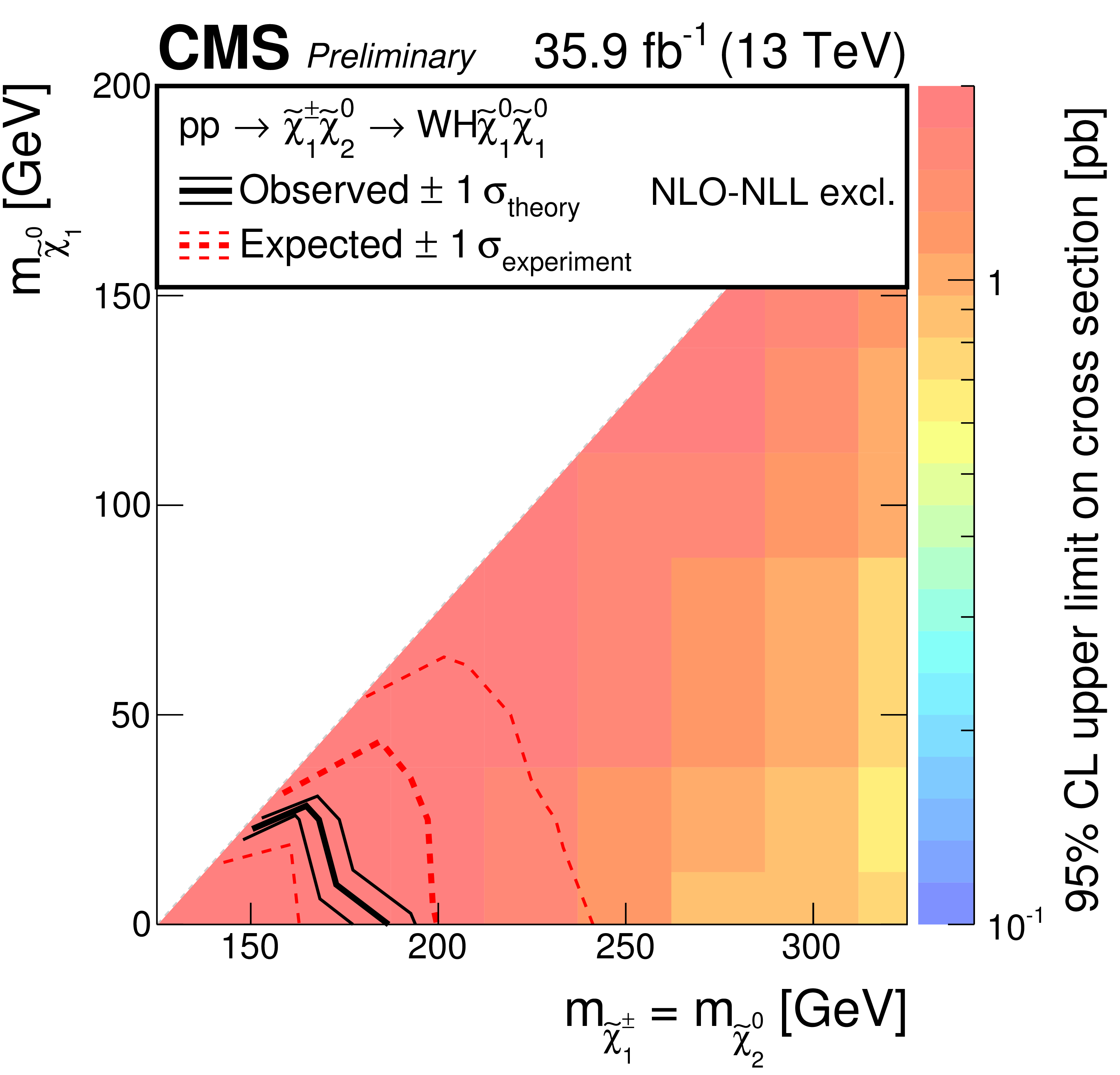} 
\caption{Interpretation for the model $\tilde{\chi}_{1}^{\pm}\tilde{\chi}_{2}^{0} \rightarrow WZ\tilde{\chi}_{1}^{0}\tilde{\chi}_{1}^{0}$ (left) and $\tilde{\chi}_{1}^{\pm}\tilde{\chi}_{2}^{0} \rightarrow WH\tilde{\chi}_{1}^{0}\tilde{\chi}_{1}^{0}$ (right).}
\label{fig:int2}
\end{figure}

No significan excess is seen. Interpretations for some of the previous mentioned models 
are shown in Figure \ref{fig:int1} and \ref{fig:int2}. Exclusion limits are set 
for these models. 

\section{Conclusions}
A search for new physics with same-sign dilepton, trilepton and four lepton
events containing up to two hadronically decaying $\tau$ leptons in proton-proton 
collisions at $\sqrt{s}$ = 13 TeV is presented, using a total luminosity of 35.9 $fb^{-1}$ 
recorded by CMS.

No significant deviation from the standard model expectations is observed. The results are
used to set limits on several simplified models of supersymmetry.


\begin{thebibliography}{99}



\bibitem{bib:SUS16039}
  [CMS Collaboration],
  ``Search for electroweak production of charginos and neutralinos in multilepton final states in pp collision data at $\sqrt{s}$ = 13 TeV'', 
  CMS Physics Analysis Summary CMS-PAS-SUS-16-039, CERN, 2017.
  
\bibitem{bib:cms}
  [CMS Collaboration],
  ``The CMS experiment at the CERN LHC'', 
  JINST 3 (2008) S08004.

\bibitem{bib:pf}
  [CMS Collaboration],
  ``Commissioning of the Particle–Flow reconstruction in Minimum–Bias and Jet Events from pp Collisions at 7 TeV'', 
  CMS Physics Analysis Summary CMS-PAS-PFT-10-002, CERN, 2010.

\end{thebibliography}
\end{document}